\documentclass[12pt]{article}
\usepackage{sc3conf}
\usepackage{amsfonts}
\usepackage{epsfig}

\newcommand{\be}{\begin{equation}}
\newcommand{\ee}{\end{equation}}
\newcommand{\bea}{\begin{eqnarray}}
\newcommand{\eea}{\end{eqnarray}}
\begin{document}
\raggedbottom
\global\arraycolsep=2pt
\title{{\small \hfill OKHEP-02-09}\\Entropy Bounds in Spherical Space}

\authors{Iver Brevik,\adref{1}
 Kimball A. Milton,\adref{2}  
  and Sergei D.  Odintsov\adref{3}}

\addresses{\1ad Division of Applied Mechanics, \\
Norwegian University of Science
and Technology, N-7491 Trondheim, Norway,
  \nextaddress \2ad Department of Physics and Astronomy\\The University of 
Oklahoma, Norman 73019 USA
  \nextaddress \3ad Tomsk State Pedagogical University, 
634041 Tomsk, Russia}

\maketitle

\begin{abstract}
 Exact calculations are given for the Casimir energy for various fields
in $R\times S^3$ geometry.  The Green's function method naturally gives
a result in a form convenient in the high-temperature limit, while the
statistical mechanical approach gives a form appropriate for low temperatures.
The equivalence of these two representations is demonstrated.  Some
discrepancies with previous work are noted. In no case, even for ${\cal N}=4$
SUSY, is the ratio of entropy to energy found to be bounded.

\end{abstract}

\section{Introduction}
The remarkable appearance of the holographic principle has fostered 
the understanding that some hitherto distant branches of
theoretical physics may have a much deeper common origin 
that was expected. One significant example of this sort is the relation,
suggested by Verlinde \cite{verlinde}
between the Cardy entropy formula \cite{cardy} and
the Friedmann equation for the 
evolution of the scale factor of the universe.
Moreover, the proposal that in the early universe there exists a
holographic bound on the cosmological entropy associated with Casimir energy
 suggests that there should be a deeper relation between Friedmann cosmology and
the Casimir effect \cite{kimbook}.
Specifically, there has been much interest in studying the entropy and energy
arising from quantum and thermal fluctuations in conformal field 
theories \cite{kl,klemm}. Whether the Verlinde bound
for the ratio of the entropy to the thermal energy
can be realized in realistic situations is a matter for specific calculations.
Previous computations have been limited to the
regime of high temperatures, so they are unable to provide definitive
results. Here we obtain exact results for various fields in the $R\times S^3$
geometry, so the issues may be more decisively addressed \cite{ap}.

\section{Conformally Coupled Scalar}
\subsection{Green's Function Method}
We can start from the formalism given in Kantowski and Milton \cite{kanmil}.
The energy is given by the imaginary part of the Green's function,
\be
U=V_3\partial^0\partial^{\prime0}\Im G(x,y;x',y')\big|_{x=x',y=y'},
\ee
where the ``external'' coordinates $x$ consist only of the time.  We introduce a
Fourier transform there
\be
G(t,y;t',y')=\int\frac{d\omega}{2\pi} e^{-i\omega(t-t')}g(y,y';\omega),
\quad
g(y,y';\omega)=\sum_{lm}\frac{Y_l^m(y)Y_l^{m*}(y')}{M_l^2/a^2-\omega^2},
\ee
with $M_l^2=(l+1)^2$ for conformal coupling on $S^3$,
so the Casimir energy is
\be
U=-\frac{i}{4\pi}\int_c d\omega\,\omega^2\sum_l\frac{D_l}{M_l^2/a^2-\omega^2},
\quad D_l=(l+1)^2,\label{ce}
\ee
where the contour $c$ encircles the poles on the positive axis in a
negative sense, and those on the negative axis in a positive sense.

Temperature dependence is incorporated by the replacement
\be
\int_c \frac{d\omega}{2\pi}\to \frac{4i}{\beta}\sum_{n=0}^\infty{}',\quad
\omega^2\to -\left(\frac{2\pi n}{\beta}\right)^2.
\label{finitet}
\ee
The prime means that the $n=0$ term is counted with half
weight.

We carry out the sum on $l$ in Eq.~(\ref{ce}) by using
the general representation
\be
\sum_{m=0}^\infty \frac1{m^2-\alpha^2}=-\frac\pi{2\alpha}\cot\pi\alpha
-\frac1{2\alpha^2}.
\label{cotan}
\ee
We then make the finite-temperature replacements (\ref{finitet}) and obtain,
after dropping the contact term arising from the constant ($\propto\zeta(-2)
=0$) 
\be
U=\frac1{a}\left(\frac{2\pi a}{\beta}\right)^4
\left[\sum_{n=0}^\infty{}'\frac{n^3}{e^{4\pi^2a n/\beta}
-1}+\frac1{240}\right],
\label{klft}
\ee
which, since the summand vanishes at $n=0$, gives only exponentially small
corrections to Stefan's law, which is the result of Kutasov and Larsen 
\cite{kl}:
\be
U\sim\frac1a\frac{(2\pi aT)^4}{240},\quad (aT\gg1).
\label{sl}
\ee

\subsection{Statistical-Mechanical Approach}
We recall the usual statistical mechanical expression for
the free energy,
\bea
F=-kT\ln Z,\quad
 \ln Z=-\sum_{n=0}^\infty(n+1)^{d-2}\ln\left(1-e^{-\beta(n+1)/a}\right),
\label{freeen}
\eea
for conformally coupled scalars in $S^{d-1}$. Here the zero-point energy,
\be
E_0=\sum_{n=0}^\infty(n+1)^{d-2}\frac{n+1}{2a}=\frac1{2a}\zeta(1-d),
\ee
has been subtracted.
The specific results for two and four dimensions are
\bea
E^{d=2}_0=-\frac1{2a}\frac12B_2=-\frac1{24a},\quad
E^{d=4}_0=-\frac1{2a}\frac14B_4=\frac1{240a}.\label{2.18}
\eea

For the temperature dependence, we differentiate the partition
function,
\be
E=U-E_0=
-\frac{d}{d\beta}\ln Z=\frac1a\sum_{n=1}^\infty\frac{n^3}{e^{2\pi n\delta}-1},
\quad\delta=\frac{\beta}{2\pi a}.
\label{smrep}
\ee
This is a very different representation from Eq.~(\ref{klft}).  Nevertheless,
from it we may obtain the same result we found above
if we use the Euler-Maclaurin sum formula.

\subsection{Relation Between Representations}
\label{sec:reps}
We have two representation for the Casimir energy, the one obtained
from the Green's function, Eq.~(\ref{klft}), and the one obtained from the
partition function, Eq.~(\ref{smrep}).  The relation between the two
can be found from the Poisson sum formula.  If the Fourier transform of
a function $b(x)$ is defined by
\be
c(\alpha)=\int_{-\infty}^\infty\frac{dx}{2\pi}e^{-i\alpha x}b(x),\quad
\mbox{then}\quad
\sum_{n=-\infty}^\infty b(n)=2\pi\sum_{n=-\infty}^\infty c(2\pi n).
\label{psf}
\ee
It is then easily seen that the energy (\ref{smrep}) is
\be
E=\frac1a\left(\frac{a}{\beta}\right)^4\Gamma(4)\sum_{n=-\infty}^\infty
\sum_{k=0}^\infty\frac1{\left(1+k-ia2\pi n/\beta\right)^4}.
\ee
If we sum this on $n$ first, we obtain the alternative expression (\ref{klft})
\bea
E=\frac1a\left(\frac{2\pi a}{\beta}\right)^4\sum_{n=1}^\infty
\frac{n^3}{e^{4\pi^2 an/\beta}-1}+\frac1{240a}\left[(2\pi aT)^4-1\right].
\label{relation}
\eea
The two representations are best 
adapted for the low- and high-temperature limits, respectively:
\bea
U=\frac1{240a}+\frac1a\sum_{n=1}^\infty\frac{n^3}{e^{2\pi n\delta}-1}
=\frac{(2\pi aT)^4}{240a}+\frac{(2\pi a T)^4}a\sum_{n=1}^\infty\frac{n^3}
{e^{2\pi n a/\delta}-1}.
\eea

\section{$d=2$ Conformal Scalar}
\label{sec:d2scalar}
Because there is a subtle issue involving zero-modes here, it is useful
to repeat the above calculation for $d=2$.  From the partition function
we immediately obtain the low-temperature representation,\footnote{Dowker
\cite{dowker} has suggested that the zero-mode contribution be retained
here.  We see no reason to include the $n=0$ term, which would in
any case lead to a violation of the Third Law of Thermodynamics
\cite{ap,elizalde}.}
\be
U=-\frac1{24a}+\frac1a \sum_{n=1}^\infty\frac{n}{e^{2\pi n\delta}-1},
\label{ltemplimd2scalar}
\ee
displaying an exponentially small correction to the zero-point energy if
$\beta\gg1$.  Again, by use of the Euler-Maclaurin sum formula we can obtain
the high-temperature limit,
\be
U\sim\frac1{24a}(2\pi a T)^2-\frac12 T,\quad (aT\gg1).
\label{htemplimd2scalar}
\ee
This again coincides with the result found in Kutasov and Larsen \cite{kl}.  
However,
the linear term in $T$ is omitted in the analysis of Klemm et al.\ \cite{klemm}
with an apparently erroneous remark that it only contributes when the
saddle-point method breaks down, for a small central charge (the number of
fields).  So their derivation of the Cardy formula  cannot
be sustained.

It is easy to reproduce this result from the Green's function method.
After the finite-temperature substitutions, the expression is
\bea
U&=&\frac1{24a\delta^2}-\frac12 T-\frac1{a\delta^2}\sum_{n=1}^\infty\frac{n}
{e^{2\pi n/\delta}-1},
\label{general2drep}
\eea
which gives the explicit exponential corrections to the high temperature
limit (\ref{htemplimd2scalar}).
The low-temperature limit displayed in Eq.~(\ref{ltemplimd2scalar}) may
be easily obtained from this by using the Euler-Maclaurin sum formula.
A proof of the equivalence of the two representations (\ref{general2drep})
and (\ref{ltemplimd2scalar}) can be carried out along the lines sketched above,
with due care for the presence of the zero-mode at $n=0$ \cite{ap}.

\section{Vector Field}

The analysis proceeds similarly to that given above.
For $S^{d-1}$ the degeneracy and eigenvalues are
\bea
D_l=\frac{2l\left(l+\frac{d}2-1\right)(l+d-2)(l+d-4)!}{(d-3)(l+1)!},\quad
M_l^2=l(l+d-2),
\eea
so for $d=4$ if we add the conformal coupling value 1 to $M_l^2$
we obtain for the Green's function mode sum
\bea
\sum_{l=0}^\infty \frac{2l(l+2)}{(l+1)^2/a^2-\omega^2}\to
-\frac1{\omega^2}+i\pi a^2\left(\omega a-\frac1{\omega a}\right)
\left(1+\frac2{e^{-2\pi i\omega a}-1}\right).
\eea
After making the finite temperature replacement, we carry out the sum on
$n$, with the result
\bea
U=\frac{(2\pi aT)^4}{120 a}-\frac{(2\pi a T)^2}{12a}+T
+\frac{2(2\pi a T)^2}{a}\sum_{n=1}^\infty\frac{n+(2\pi a T)^2
n^3}{e^{4\pi^2aTn}-1},
\label{htexv}
\eea
where the $T$ term comes from the $n=0$ term in the sum. 
 Since the remaining
sum is exponentially small in the large $T$ limit, this form is well-adapted
for high temperature. 
(The $T^4$ and $T^2$ terms are as given in Kutasov and Larsen \cite{kl}.)
 However, it is exact, and by using the Euler-Maclaurin
sum formula it yields the low temperature limit,
$U\sim\frac{11}{120a}$, $aT\ll1$,
up to exponentially small corrections.  The latter may be directly inferred
from the partition function,
\be
\ln Z=-\sum_{l=1}^\infty 2l(l+2)\ln\left(1-e^{-\beta(l+1)/a}\right).
\ee
  By taking the negative derivative of this with respect to $\beta$ we
obtain the alternative representation
\be
U=\frac{11}{120a}+\frac2a\sum_{l=1}^\infty\frac{l(l^2-1)}{e^{\beta l/a}-1}.
\label{ltexv}
\ee
The  exact equivalence of the two expressions (\ref{ltexv}) and
 (\ref{htexv}) again is demonstrated either by
 use of the Euler-Maclaurin sum formula, or by the Poisson sum formula,.

\section{Weyl Fermions} 
Here, the degeneracies and eigenvalues are
$D_l=2(l+2)(l+1)$, $M_l^2=(l+3/2)^2$,
so including the minus sign associated with a fermionic trace,
and the antiperiodicity of the fermionic thermal Green's functions,
we have the following expression for the energy,
\bea
U&=&\frac1a\Bigg\{\frac7{960}\delta^{-4}-\frac1{96}\delta^{-2}
-\frac14\sum_{n=0}^\infty\left[(2n+1)^3\delta^{-4}+(2n+1)\delta^{-1}
\right]\nonumber\\
&&\quad\times\left(\frac2{e^{2\pi(2n+1)/\delta}-1}-\frac1{e^{\pi(2n+1)/\delta}-1}
\right)\Bigg\}.
\label{htexf}
\eea
The low-temperature limit (the zero-point energy) may be obtained from
this by the Euler-Maclaurin formula, and the exponential
corrections in that limit may be obtained directly from the 
partition function,
\be
\ln Z=\sum_{n=1}^\infty 2n(n+1)\ln\left(1+e^{-\beta(2n+1)/2a}\right).
\ee
That is
\be
U=\frac1a\left[\frac{17}{960}+\sum_{n=1}^\infty \frac{n(n+1)(2n+1)}
{e^{\beta(2n+1)/2a}+1}\right].
\label{ltexf}
\ee
The equivalence between Eqs.~(\ref{htexf}) and (\ref{ltexf}) may be
demonstrated as above.

\section{Entropy Bounds}
\label{sec:VI}
From the above results, thermodynamic information may extracted in terms of
the free energy,
in terms of which the energy and the entropy may be extracted:
\be
E\equiv U-E_0=-\frac\partial{\partial \beta}\ln Z=\frac\partial{\partial\delta}
\delta F,\quad
S=2\pi a\delta^2\frac{\partial}{\partial\delta}F=\beta(E-F).
\ee

\subsection{Two-dimensional scalar}
Klemm et al.\ \cite{klemm} ignore the linear $T$ term in the energy,
and so have
\bea
E=\frac1{24a}(\delta^{-2}+1),\quad
F=-\frac1{24a}(\delta^{-2}-1),\quad
S=\frac\pi6\delta^{-1}.
\eea
These imply the Verlinde-Cardy formula, and the entropy bound,
\be
S=4\pi a\sqrt{E_0(E+E_0)},
\quad
\frac{S}{2\pi a E}=2\frac{\delta}{\delta^2+1}\le1.
\ee
However, this result is not meaningful as it stands.  Even in the 
high-tempera\-ture limit we must add the term linear in temperature to the 
energy, which implies instead from Eq.~(\ref{htemplimd2scalar}), for
$\delta\ll1$, that
\bea
E=\frac{\delta^{-2}+1}{24a}-\frac1{4\pi a\delta},\,
F=-\frac{\delta^{-2}-1}{24a}-\frac{\ln\delta}{4\pi a\delta},
\, S=\frac\pi{6\delta}+\frac{(\ln\delta-1)}2.
\eea
The ratio of $S$ to $E$ is then unbounded as $\delta\to\infty$.  Yet this
takes us to the low-temperature regime, where we must use the leading
exponential corrections,
\bea
E\sim\frac1a e^{-\beta/a},\quad
F\sim-\frac1\beta e^{-\beta/a},\quad
S\sim\frac\beta{a} e^{-\beta/a},\quad (\beta\gg1)\label{lowtemp2dentropy}
\eea
so the entropy-energy ratio is
\be
\frac{S}{2\pi a E}=\delta,\quad (\delta\gg1).
\ee
It is apparent that this latter result is universal because the energy
always dominates the free energy in the low temperature regime.

\subsection{Entropy Bounds in Four Dimensions}
In the following we will consider cases with $N_s$ conformal scalars,
$N_v$ vectors, and $N_f$ Weyl fermions. 
In the high-temperature regime we may write the free energy, energy, and
entropy as\footnote{What is called the Cardy formula is simply
the observation that the leading behavior of $S$ is the geometric
mean of the leading and subleading terms in $E$.  The term ``Casimir energy''
for the latter is misleading in other than $1+1$ dimensions.
The entire energy $U$ is due to quantum and thermal fluctuations, so it all
should properly be reckoned as Casimir energy.}
\bea
F&\sim&-\frac1a[a_4\delta^{-4}+a_2\delta^{-2}+a_1\delta^{-1}\ln\delta+a_0],
\\
E&\sim&\frac1a[3a_4\delta^{-4}+a_2\delta^{-2}-a_0-a_1\delta^{-1}],\label{form}\\
S&\sim&2\pi[4a_4\delta^{-3}+2a_2\delta^{-1}-a_1(1-\ln\delta)].
\eea
Here the coefficients were determined in the previous sections to be
[see Eqs.\ (\ref{relation}), (\ref{htexv}), and (\ref{htexf})]
\bea
a_4=\frac{N_s}{720}+\frac{N_v}{360}+\frac{7N_f}{2880},\,
a_2=-\frac{N_v}{12}-\frac{N_f}{96},\,
a_0=3a_4-a_2,\,
a_1=-\frac{N_v}{2\pi}.
\eea
Even ignoring the $a_1$ term, Klemm et al.~\cite{klemm} note that no
entropy bound is possible, unless special choices are made for the field
multiplicities.  For the ${\cal N}=4$ case ($N_s=6$, $N_v=1$, $N_f=4$)
 the entropy-energy ratio becomes
\be
\frac{S}{2\pi a E}=\frac{1-\ln\delta+\frac{\pi}6\delta^{-3}(1-3\delta^2)}
{\delta^{-1}+\frac{\pi}8\delta^{-4}(1+\delta^2)(1-3\delta^2)}.
\label{enenratio}
\ee
If the $a_1$ terms here were omitted, the zero in both the energy and
entropy at $\delta^2=1/3$ would cancel, and we would have the limit
given by Klemm et al.~\cite{klemm}:
\be
\frac{S}{2\pi a E}=\frac43\frac\delta{1+\delta^2}\le\frac43
\label{klemmratio}
\ee
in the high temperature regime.  But $a_1\ne0$, and the ratio (\ref{enenratio})
diverges as $\delta\to\infty$.  Of course that limit is the low-temperature one,
but the argument given above then applies and shows that
\be
\frac{S}{2\pi a E}\sim\delta,\quad (\delta\to\infty).
\ee
Although in this limit both the entropy and the subtracted energy are
exponentially small, their ratio is unbounded.

It should noted that we are not in formal disagreement with previous
studies.  The interest there was restricted to high
temperature, which is presumably all that is relevant to 
nearly the entire history
of the universe.\footnote{Before or
after photon decoupling, but after inflation,
the value of $aT$ stays nearly constant,
essentially reflecting entropy conservation.  That value
is far into the high temperature regime, the present value of $\delta$
being $\delta_0\sim 10^{-30}$.  Insofar as it is permissible to speak
of temperature during inflation, $aT$ is also constant then, but of a
much smaller value, which value increases dramatically during reheating.} 
  In that case, only the leading terms in $1/\delta$ are
relevant, and the ratio of entropy to energy is always of order 
$\delta\ll1$.
It is not surprising that such results as Eq.~(\ref{klemmratio}) are
an unreliable guide to the moderate and low temperature regimes,
which might be relevant in the very earliest 
(pre-inflationary) stages of the universe.

Another point, which is more closely connected with physics, is that it is 
permissible to make use of the thermodynamical formalism for fluctuating 
quasi-classical systems only when the temperature $T$ is sufficiently high
\cite{das}, that is,
$\delta \ll 1$.
 It seems that we must be careful in not assigning too much 
physical significance to the subleading corrections.

\section{Conclusion}

Our main qualitative result is that entropy/energy bounds
should be relevant only in the
ultra-high temperature limit, which
applies to the universe after inflation.
With the decrease of temperature the bound becomes much less 
reliable, until at low temperature ($aT\ll1$),
which might be the case in the very early universe,
the entropy dominates the energy.
This effect occurs already for conformal matter.
The situation for non-conformal matter is much more 
complicated.  Hence, it is unclear if entropy/energy bounds should exist 
at all even for high temperature. \normalcolor
This question will be discussed 
elsewhere.

\section*{Acknowledgement}
KAM thanks the US Department of Energy for partial financial support of his
research.  He is grateful for the invitations of the Third International
Sakharov Conference on Physics for the invitation to present this work there.

\end{document}